\documentclass[preprint2]{aastex63}

\begin{document}

\title{Masses and densities of dwarf planet satellites measured with ALMA}

\author[0000-0002-8255-0545]{Michael E. Brown}
\affiliation{Division of Geological and Planetary Sciences\\
California Institute of Technology\\
Pasadena, CA 91125, USA}
\author[0000-0002-5344-820X]{Bryan J. Butler}
\affiliation{National Radio Astronomy Observatory, Socorro NM 87801 (U.S.A.)}

\begin{abstract}
We have used the Atacama Large Millimeter Array (ALMA) to measure precise 
absolute astrometric 
positions and detect the 
astrometric wobble of dwarf planet Orcus and its satellite Vanth over a complete orbit. We also
place upper limits to the astrometric
wobble induced by Dysnomia on dwarf planet Eris 
around its orbit.
From the Vanth-Orcus barycentric motion, we
find a Vanth-Orcus mass ratio of 0.16$\pm$0.02 --
the highest of any known planet or dwarf planet.  
This large ratio is consistent with the hypothesis that Vanth is a largely-intact
impactor from a giant collision in the system, 
and that the system has likely evolved to a double synchronous state.  
We find only an upper limit of the barycenter motion of Eris, 
which implies a one sigma upper limit to the Dysnomia-Eris mass ratio of 0.0085, close 
to the modeled transition region between giant impact generated 
satellites which are largely intact remnants of the original impactor and
those which form out of
reaccreted disk material left over post-impact.  
The low albedo of Dysnomia leads us to marginally favor the intact impactor scenario.  We find that Dysnomia has density of $<1.2$ g cm$^{-3}$, significantly lower than the 2.4 g cm$^{-3}$ of
Eris.
\end{abstract}

\section{Introduction}
Satellites are ubiquitous around dwarf planets in the Kuiper belt, with satellites
known around at least 8 of the 10 largest dwarf planets. Smaller
Kuiper belt objects (KBOs) are often found as nearly equal-sized binaries of similar color on eccentric 
orbits \citep{2020tnss.book..201N}, and their
orbital properties have lead to the hypothesis that they are formed from direct collapse via gravitational
instabilites \citep{2019NatAs...3..808N}. Dwarf 
planet satellites, in contrast, are frequently found to be significantly smaller than their primaries and are often
found on circular or near-circular orbits, suggesting
a separate formation mechanism for these systems
\citep{2006ApJ...639L..43B}. 

Giant impacts have long been discussed as a likely 
formation mechanism for dwarf
planets satellites. A giant impact origin for 
Pluto's satellite Charon was proposed by \citet{1989ApJ...344L..41M},
and \citet{2005Sci...307..546C} showed that a relatively large and high
density satellite such as Charon could indeed
be formed through an oblique giant impact, where the
impactor is captured largely intact after the collision.

\citet{2019NatAs...3..802A} modeled a wider range of dwarf planet 
collisions and found that most of the known dwarf planet 
satellite systems appear consistent with formation via a giant
impact that occurred at close to the escape velocity,
followed by the retention of the largest fragment from the
impactor. Depending on the impact angle, this largest
captured fragment can range from a nearly-intact
original impactor (as in the case of Pluto-Charon), to a
low density icy fragment containing only a small fraction of the
mass of the original impactor. The requirement that the
impact occur at close to the escape velocity is a strong
indicator that these collisions happened early in
solar system history before dynamical excitation 
of the Kuiper belt.

Two critical parameters for understanding the formation of
dwarf planet satellites and for testing models such as 
these giant impact scenarios are the satellite-primary
mass ratio and the density of the satellite compared to
the primary. Such parameters are known for only two
systems -- Pluto-Charon and Haumea-Hi'iaka-Namaka --
which span a wide range of parameter space.

The Charon-Pluto mass ratio is 0.12 -- the largest
measured for any planet or dwarf planet -- and was determined
by detecting the motion of Pluto and Charon around their
common barycenter
\citep{1994Icar..108..186Y, 1996AJ....111.1368N, 1997Icar..126..362F, 2003Icar..164..254O, 2006AJ....132..290B}.
Densities of Pluto and Charon are 1.85 and 1.7 g cm$^{-3}$, respectively,
approximately equal and typical for objects of their sizes.

In contrast, the Haumea system has a (total) satellite-primary mass ratio
of 0.0045, where the mass of the satellites was determined by
detecting their mutual perturbations \citep{2009AJ....137.4766R}. The sizes
of the satellites have not been measured directly, but with
their low masses and surface spectra that resemble
pure water ice \citep{2006ApJ...640L..87B, 2009ApJ...695L...1F}, it appears likely that
the objects are icy fragments with densities of $\lesssim$ 1
g cm$^{-3}$, in marked contrast to the 1.8-2.0 g cm$^{-3}$ density of Haumea
\citep{2017Natur.550..219O,2019ApJ...877...41D}.

For both of these systems, measurement of their mass ratios
relies on the presence of multiple satellites.
Measuring the mass of dwarf planet
satellites which have no additional satellite companions
is more difficult. 
The only plausible method to 
measure satellite mass is through the detection
of the barycenter 
motion of the primary determined against an
absolute astrometric background.
Barycenter motion has not been possible
to measure with high resolution optical or infrared
imaging to date owing to insufficient astrometric references
in the same field of view as the target moves against 
the background stars.

Unlike high resolution
optical or infrared imaging, radio interferometric
observations routinely measure positions in an absolute
astrometric reference frame during the process 
of phase calibration, usually using 
extragalactic radio sources which define the
standard celestial reference frame \citep{2009_Ma,2019_Petrov}.
There is a well-established history of obtaining
precise positions of sources over time from such observations
\citep{2014_Reid}.  
We use the extraordinary ability of ALMA to provide absolute
astrometry to allow us to search for
the barycentric motion of Eris and Orcus caused by
their satellites.

Eris is the most massive dwarf planet known and has
a single known satellite in a circular orbit
at a distance of $a/R_p\approx 32$ from the primary, where $a$
is the radius of the orbit and $R_p$ is the radius of the 
primary. ALMA observations
give a diameter of Dysnomia of $700\pm115$ km \citep[][hereafter BB18]{2018_Brown}, making
it the second largest known satellite of a dwarf planet, and
suggesting that the Dysnomia-Eris mass ratio could be
anywhere from below 0.01 to 0.03, depending on whether or
not Dysnomia has a density below 1.0 g cm$^{-3}$ -- as 
appears typical for KBOs of this size -- or has a density
more similar to the 2.4 g cm$^{-3}$ value derived
for Eris \citep{2011_Sicardy}. The Eris-Dysnomia system appears
in a regime different from either that of Pluto or 
that of Haumea.

The Orcus-Vanth system, in contrast, appears like
a scaled-down version of the Pluto-Charon system. Vanth orbits
on a circular orbit at a distance of $a/R_p\approx 20$
from the
primary (compared to $\sim$16.5 for Pluto-Charon), 
and ALMA observations have measured diameters of
910$^{+50}_{-40}$ and $475\pm75$ km for Orcus and Vanth (BB18), a size ratio of
1.6$\pm$0.3 (comparable to the value of 2.0 for Pluto-Charon). { The ALMA-derived 
effective diameter of Vanth is consistent with that measured from a stellar
occultation and an assumed spherical shape of  $443\pm10$ km \citep{2019Icar..319..657S}, though we will conservatively use the ALMA result for consistency between
Orcus and Vanth}. Assuming identical
densities, the Vanth-Orcus mass ratio would be 0.142$\pm$0.02, a value
even higher than the 0.12 of Pluto-Charon.

These observations will expand the range of dwarf planet
and satellite sizes, ratios, and orbital distances with
fully characterized systems,
allowing us to continue to explore formation mechanisms
for these ubiquitous satellite systems.

\section*{Observations and Data Reduction}

The Orcus-Vanth system was observed 4 times in Oct/Nov 2016.  The Eris-Dysnomia
system was observed 3 times in Nov/Dec 2015.
A complete description of the observations, including the method for obtaining
final flux densities, is contained in BB18.  We describe
here only the further steps in the data reduction required to obtain the
astrometric positions and errors.

To measure the positions of the detected objects, we perform direct
fits to the { interferometric}
visibilities.  In the case of Orcus and Vanth, we use a model
of two point sources, with initial position estimates given by Gaussian
fits to the images.  In the case of Eris, we use a model of a slightly
limb-darkened disk, with diameter 1163 km \citep{2011_Sicardy}, also 
with initial position estimate given by Gaussian fits to the images.

We assume three sources of astrometric error: formal
fitting uncertainties (``ff error''), systematic fitting uncertainties (``sf error''), and overall
celestial frame uncertainties (``cf error''). Formal fitting
uncertainties are simply the errors returned in the visibility fits.  We 
estimate the systematic fitting uncertainties by differencing the
positions returned by the visibility fitting with those returned by the
Gaussian image fits.  Our methodology for finding the celestial frame
error is described below.  We then take the total error in either direction
(right ascension or declination) as the root sum squared (RSS) of
those three values.

The most difficult error term to determine is the overall celestial
frame uncertainty.  Fortunately, astrometric observations with ALMA
always contain at least two ``check sources,'' which are sources near
the science target source which have well-determined positions.
Additionally, we used a primary phase calibrator which also has a
well-determined position for both sets of observations.  The positions
of these sources are either taken from the International Celestial
Reference Frame (ICRF), or the Radio Fundamental Catalog (RFC)
\citep{2009_Ma,2019_Petrov}.  For the Eris observations, the primary
phase calibrator was J0125-0005 (ICRF; 5.2 degrees distant) and the two check sources were
J0141-0928 (ICRF; 6.5$^\circ$) and J0115-0127 (RFC; 7.0$^\circ$).  Unfortunately, J0141-0928 was
sufficiently far from Eris that phases did not
transfer from the primary phase calibrator well, so we did not use it.
For the Orcus and Vanth observations, the primary phase calibrator was
J1048-1909 (ICRF; 13.0$^\circ$), and the two check sources were J1022-1037 (RFC; 2.2$^\circ$) and
J0942-0759 (RFC; 6.0$^\circ$).  Both were sufficiently close to Orcus and Vanth that
they could be used.  On each date, we made an image of the check
sources, and did a Gaussian fit to find the offsets of those sources
from their expected positions (which should be at the phase center, or
the image center, if everything worked perfectly).  For Eris we took
the offset of J0115-0127 as the frame uncertainty; for Orcus and Vanth
we took the average of the offsets of J1022-1037 and J0942-0759 as the
frame uncertainty.

Table \ref{astrometry} shows the final positions, all
contributions to the error, and the final error.  For Eris, the errors
are of order a few mas; for Orcus and Vanth they are a factor of
roughly 2-3 higher on two of the days, but much larger for the two others.  The
observation on October 13 was impacted by poor observing conditions;
that on November 7 was taken when antennas had been moved into a more
compact configuration.  Normally, neither of these would happen for ALMA
observations, as they are scheduled purposefully when conditions and
resolution is appropriate to the proposed science.  However, for these
observations, there were very tight time constraints - both because
observations must be taken at separated orbital phases (which means a
few days apart), and because the time spent in each configuration is
limited.  If all observations had been taken under ideal circumstances,
the errors would almost certainly all be as they are for Eris, a few
masec.  This uncertainty agrees with the expected astrometric accuracy of ALMA
\citep{2018_Warmels}.


Figures 1 and 2 show the astrometric measurements of Orcus and Vanth, 
and of Eris, respectively. Dysnomia is too faint to be detected in the individual 
images.
The barycentric motion of the Orcus-Vanth system can clearly be seen in the data. 
For Eris some barycenter deviation is apparent, but it is much smaller and less regular than that of Orcus-Vanth.
\begin{figure}

    \plotone{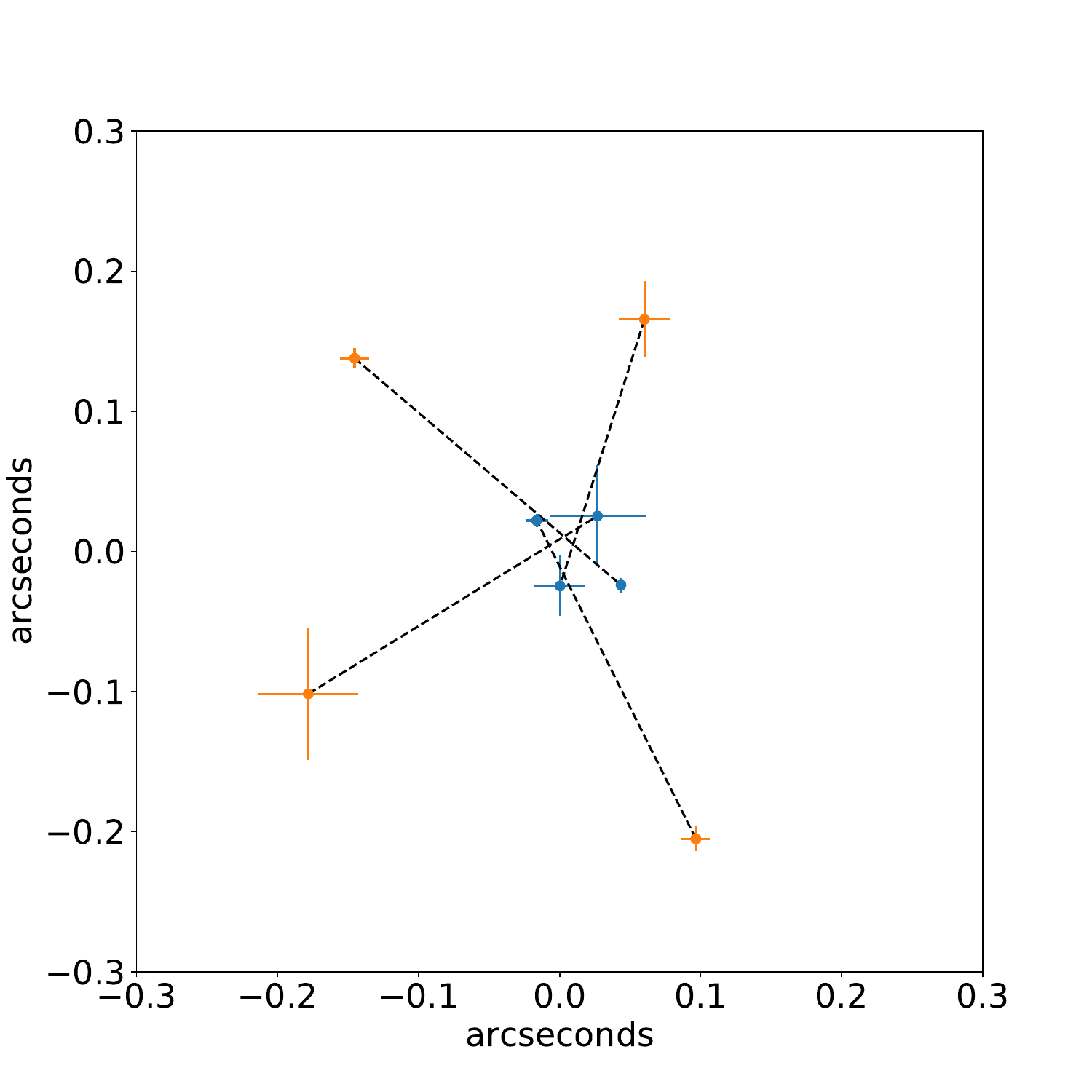}
    \caption{Astrometric positions of Orcus (blue) and Vanth (orange) at four positions in 
    the orbit of Vanth, referenced to an absolute astrometric reference, corrected
    for the predicted motion of the Orcus system through the sky, and with the mean offset from
    the ephemeris prediction removed. The dashed lines connect the positions of Orcus and Vanth
    measured simultaneously. The barycentric motion of Orcus about the center-of-mass can
    be seen in the motion of Orcus opposite that of Vanth.}
    \label{fig:ov}
\end{figure}

\begin{figure}

    \plotone{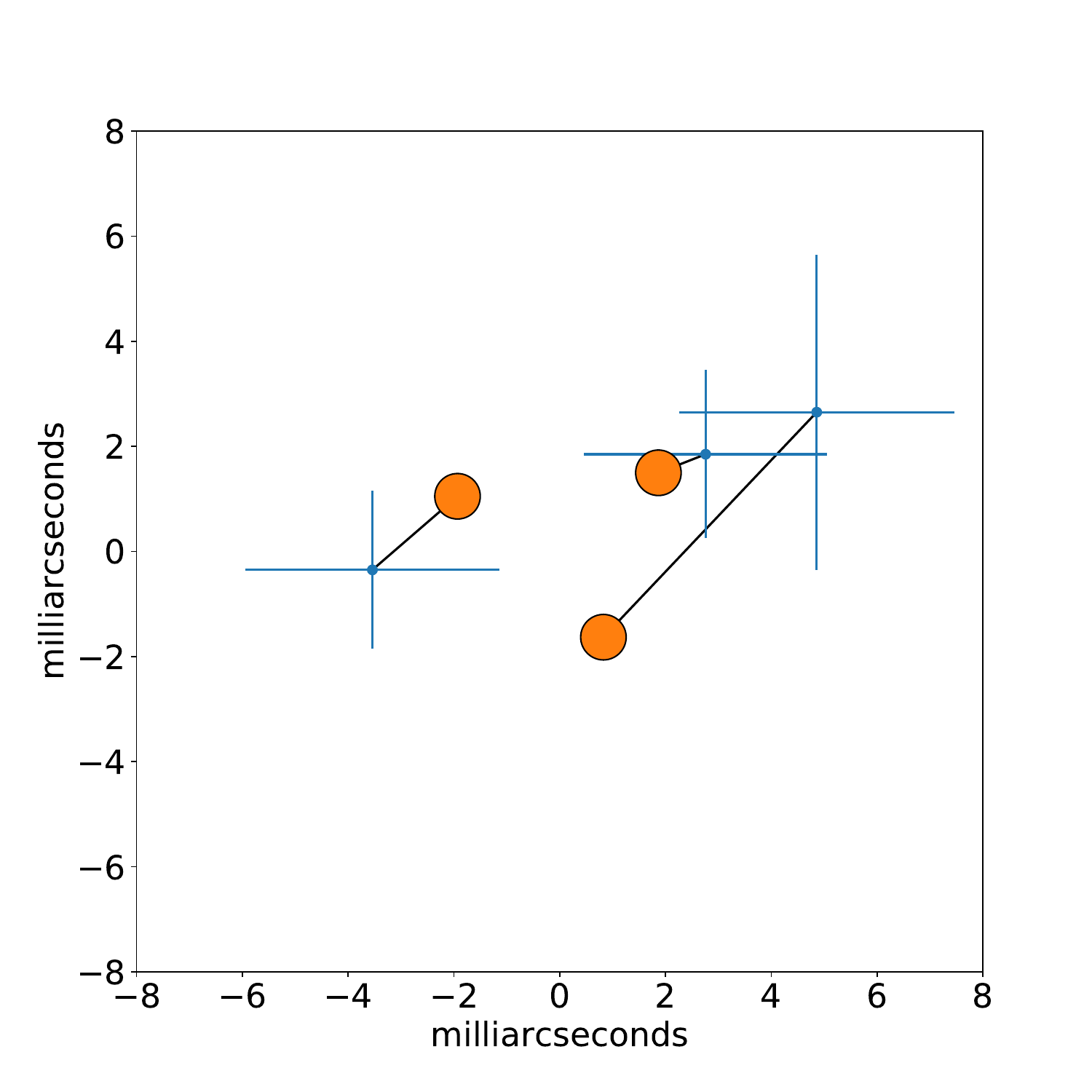}
    \caption{Astrometric positons of Eris at
    three separate positions in the orbit of 
    Dysnomia, referenced to an absolute astrometric frame, corrected for the
    predicted motion of the Eris system through the sky,
    and with the mean offset from the ephemeris prediction removed, show as blue points
    with uncertainties. The orange points show
    the maximum likelihood fits to the 
    barycenter motion, which is only detected
    at the 1.5$\sigma$ level. { The barycentric motion is
    much smaller than Eris itself, which subtends 33 mas.
    Dysnomia is between 330 and 540 mas away during these
    observations.}}
    \label{fig:ed}
\end{figure}
\begin{figure}
    \centering
    \includegraphics[scale=.5]{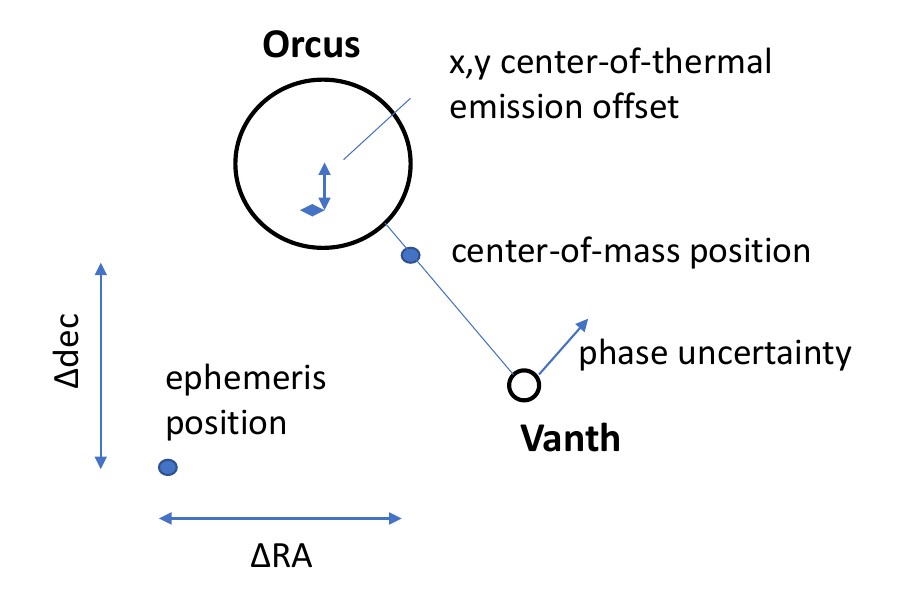}
    \caption{ The six free parameters to be fit in the Orcus-Vanth maximum likelihood model
    are a phase offset of Vanth in its orbit (the other orbital parameters are 
    unconstrained by the ALMA data), the Orcus-Vanth mass ratio, which 
    uniquely defines r.a. and dec offsets to Orcus and Vanth from the center of mass,
    an r.a. and dec offset from the predicted ephemeris position to the actual ephemeris
    position, accounting for the poorly known position of the system (and assumed
    to be constant over the short time scale of the observations), and a potential
    r.a. and dec offset between the center of light of Orcus (measured during the
    HST observations defining the orbit of Vanth) and the center of thermal
    emission (measured by ALMA). Note that this final parameter, which could
    have been significant, is found to be consistent with zero.}
    \label{fig:my_label}
\end{figure}

\section{Analysis}
\subsection{Orcus-Vanth}
To determine the center of mass of the Orcus-Vanth system, 
we fit the deviations of Orcus 
and Vanth from the expected system ephemeris position to a 6-parameter likelihood model 
that consists of the following (Fig. 3): (1) an orbital phase offset to the 
position of Vanth from that determined in
BB18 to allow for the uncertainty 
in the decade-old orbital phase measurement from the
Hubble Space Telescope (HST) {(note that 
uncertainties in the other orbital parameters are
significantly smaller and would not affect the results here)}; (2) a satellite-primary mass ratio;
(3-4) a constant ephemeris offset from
the expected position of the pair to the center of mass; and (5-6) a constant 
offset between the center-of-light observed with ALMA and the center-of-light measured at optical wavelength. This last parameter
could be significant if, for example, enhanced thermal emission was coming from a 
sunward-facing pole offset from the projected center of the body. At the 
projected size of Orcus of nearly 28 mas, this parameter could be large.

We evaluate this 6-parameter model using the Markhov Chain Monte Carlo method implemented
in {\it emcee} \citep{2019JOSS....4.1864F}.
After an initial burn-in, we collect 100000 samples of the Markhov chain (MCMC)
for analysis. The distributions of the parameters are nearly Gaussian, so we report the median and the
15.9\% and 84.1\% intervals as the results and uncertainties. 
Table 2 gives the retrieved values of all parameters. 
Figure 4 shows the predicted positions of Orcus and Vanth using the median of all parameters. We find that the center-of-light measured with ALMA and 
at optical wavelengths is the same (measured offsets of $2\pm3$ mas both
east-west and north-south compared to the 14 mas projected radius of
Orcus). Such a measurement, along with the 
the nearly 
pole-on orbit of Vanth and the lack of a significant light curve for
Orcus \citep{2016Ap&SS.361..212G}, suggests that we are viewing Orcus pole on. 
\begin{figure}
    \centering
    \plotone{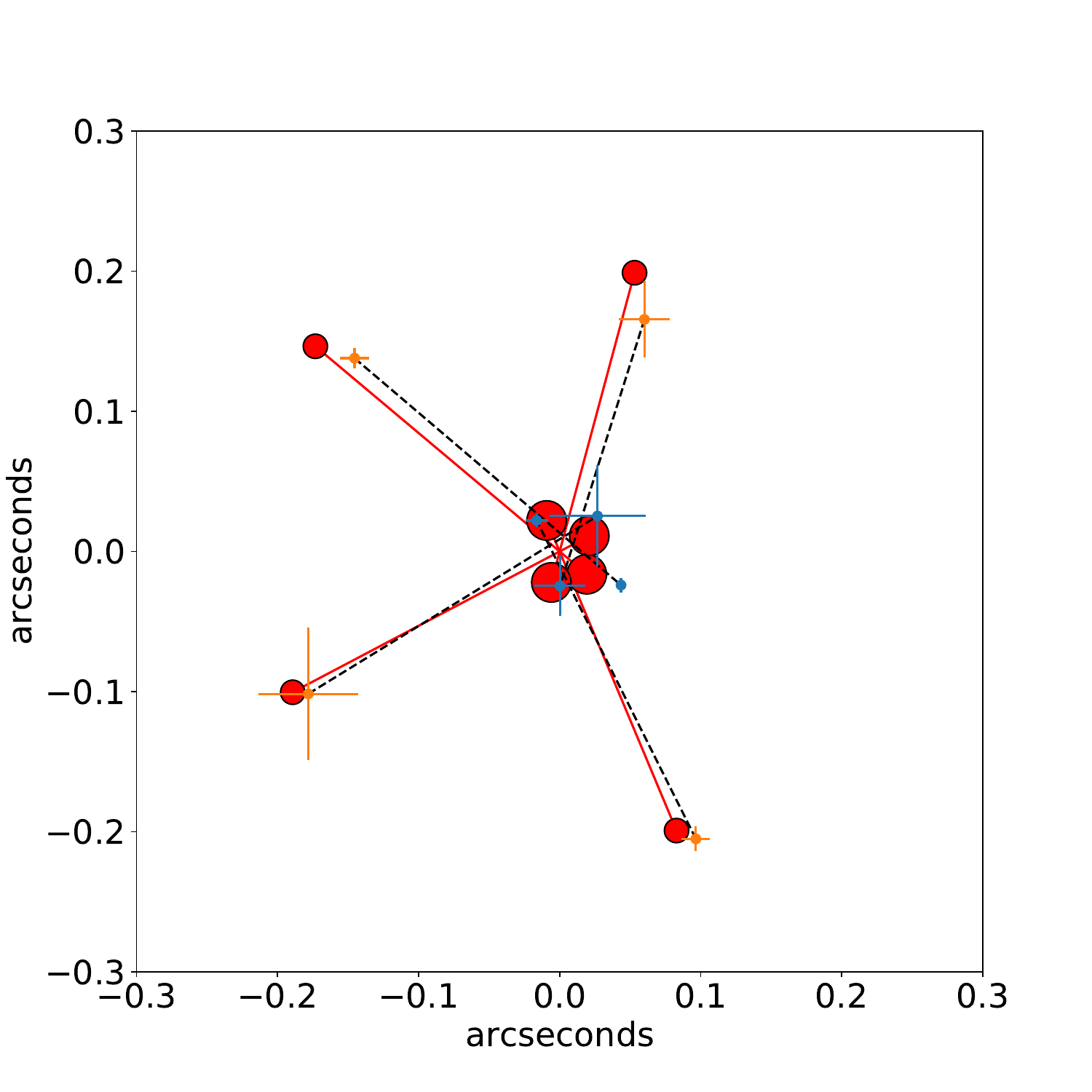}
    \caption{The best-fit model to the barycentric motion of Orcus and Vanth, compared to
    the astrometric measurements. The astrometric measurements are
    the points with uncertainties reproduced from Figure 1, while the
    model shows the predicted positions of Orcus and of Vanth for each
    observation (again, with observations at the same epoch connected
    by a dashed line). The model provides an excellent fit to the data, with a reduced $\chi^2$ of 0.95 and most of the
    data points within $\sim1 \sigma$ of their predicted positions and the barycentric offset
    angular sizes of Orcus and of Vanth. Note that, like the Pluto-Charon system, the
    barycenter of the Orcus-Vanth system, seen at the intersection of the
    4 orange lines, lies exterior to the primary.}
    \label{fig:my_label}
\end{figure}

\begin{table}[]
\renewcommand\thetable{2}
    \centering
    \begin{tabular}{l c c }
         &  Orcus/Vanth & Eris/Dysnomia\\
       { Measured values}  &  \\
       $\Delta$RA (mas)& 15$\pm$4 & $-87\pm1$ \\
       $\Delta$dec (mas) & 3$\pm$5& $-123\pm1$\\
       phase offset (deg) & 0.2$\pm$1.7& - \\
       mass ratio& $0.16^{+0.02}_{-0.01}$ &$0.0050\pm0.0035$ \\
       x offset (mas) & 1$\pm$4 &-\\
       y offset (mas) & 1$\pm$4 &-\\
       \\
       { Derived values} \\
       sat. mass (kg) & $8.7\pm0.8\times 10^{19}$& $<1.4 \times 10^{20}$ \\ 
       sat. density (g cm$^{-3}$) & 1.5$^{+1.0}_{-0.5}$ & $<1.2$ \\
       mean anomaly (deg) & 54.2$\pm$1.7 & -\\
       epoch (JD; defined) & 2457699.93569& -\\
       
    \end{tabular}
    \caption{ Parameters of the primary-satellite systems}
    \label{tab:my_label}
\end{table}

The phase of Vanth is advanced by 4.3 degrees from the prediction using the earlier orbit.  { An updated mean anomaly and epoch for the satellite is given in Table 2.} The center-of-mass of the Orcus-Vanth
system is situated 0.137$\pm$0.013 of the way towards Vanth, outside of the body of Orcus.

This 10$\sigma$ detection of barycenter motion
directly shows that Vanth contains 13.7$\pm$1.3\% of
the mass of the system, for a Vanth-Orcus mass ratio of 0.16$\pm$0.02. 
For diameters of Orcus and Vanth of 910$^{+50}_{-40}$ and 475$\pm$75km and a system mass of $6.32\times10^{20}$ kg, 
the densities of Orcus and Vanth are $1.4\pm0.2$ and $1.5^{+1.0}_{-0.5}$ g cm$^{-3}$, respectively.

\subsection{Eris-Dysnomia}
Dysnomia is not detected in the individual images, so we must determine the center-of-mass using 
only the ephemeris deviations of
the position of Eris itself. In this case, ephemeris offset and center-of-light offset are
degenerate, so we solve only for their combination, yielding no
useful information. We use the Dysnomia ephemeris from BB18, which has uncertainties
of only 2 mas at the epoch of observation. Our model thus only has 3 parameters, the barycenter offset
and the two-dimensional ephemeris offset. We again evaluate this model with an MCMC analysis. 
Figure 2 shows the fitted model to the data. We find a 1.5$\sigma$ detection of a barycenter
motion of Eris corresponding to a mass ratio of
0.0050$\pm$.0035, or a barycenter motion of approximately
$\pm$2 mas. The predicted positions of Eris for
the maximum likelihood solution to this model can 
be seen in Figure { 2}. The predictions are moderately 
consistent with the data, with a reduced $\chi^2$ value of 1.8. 
While this barycentric motion may represent a true
detection, we chose to instead report the derived
mass ratio as a 1$\sigma$ upper limit of 0.0084 also corresponding
to a 3$\sigma$ upper limit of 0.015. Dysnomia is a small fraction of the mass of Eris;
for a 
system mass of 1.65$\times10^{22}$ kg \citep{2021Icar..35514130H}
the 1 $\sigma$ upper limit to the mass of Dysnomia is 
$1.4\times 10^{20}$ kg.

\subsection{The size and density of Dysnomia}
\begin{figure}
\epsscale{1.8}
\plotone{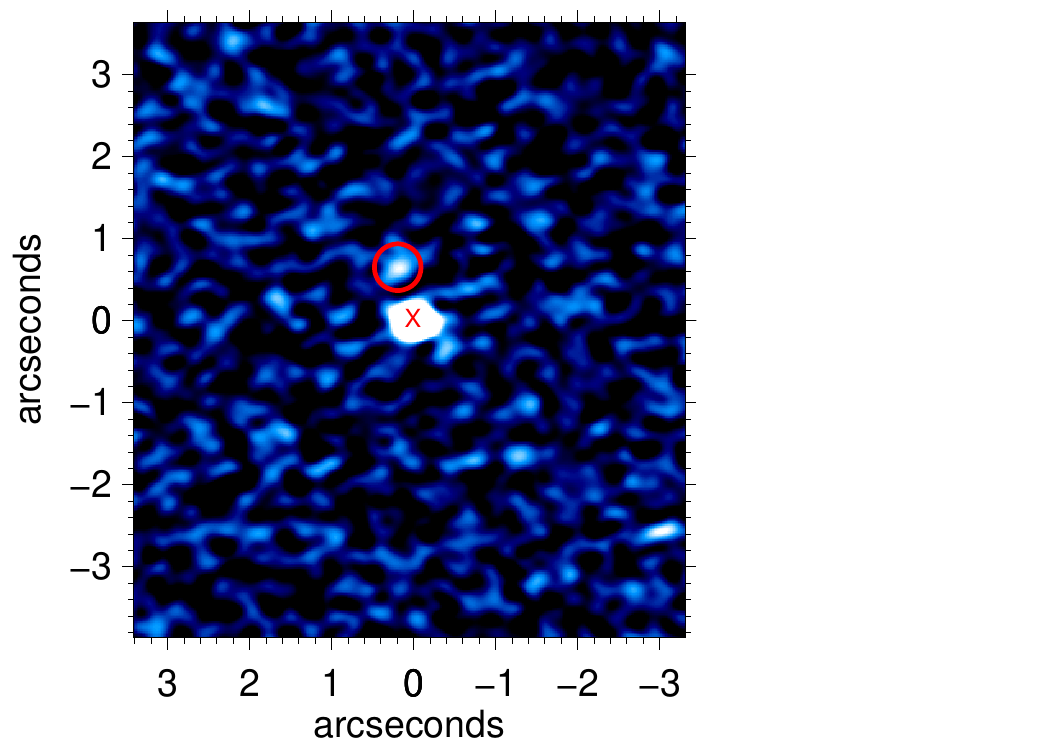}
\caption{
Image of Eris and Dysnomia from ALMA Band 6 observations on 11 October 2018. The
measured position of Eris is marked with a cross,
while the predicted position of Dysnomia with-respect-to the measured position of Eris is circled in red. Dysnomia is 
the second-brightest source in the field and appears at its predicted position
as a 6.5 $\sigma$ detection.
}
\label{fig:dysnomia}
\end{figure}

\begin{figure}
\epsscale{1.2}
\plotone{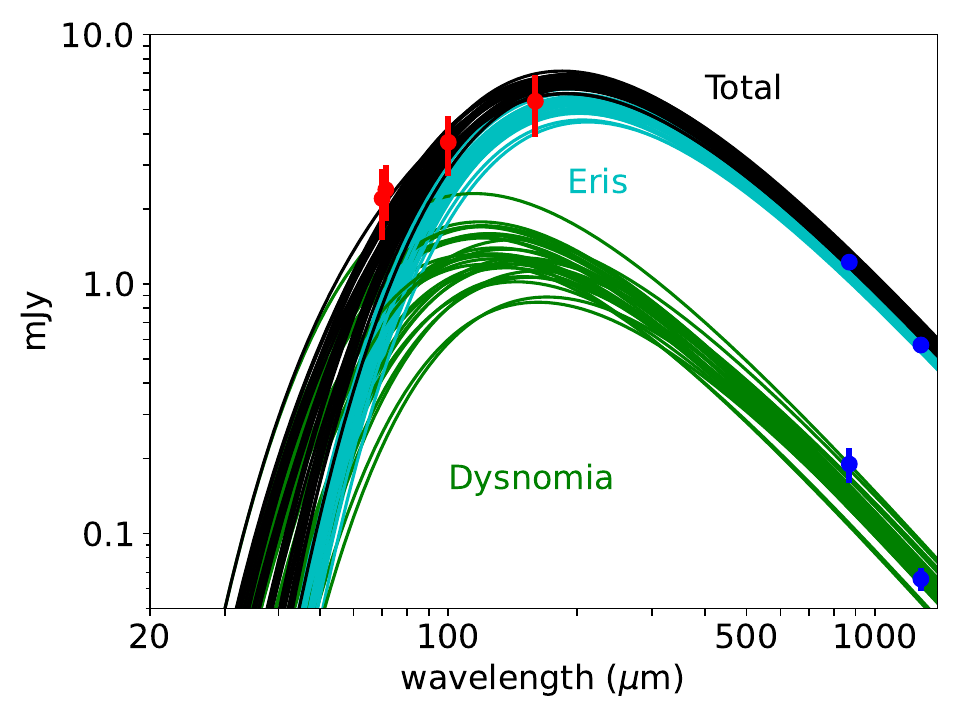}
\caption{
Unresolved flux densities measured from the Spitzer and Herschel observatories (red) along with resolved measurements from ALMA (black),
from BB18, with the current data points added at 1280 $\mu$m. Random samples
of the MCMC model of BB18 with the new flux density measurements added
are also shown.
}
\label{fig:thermal_model}
\end{figure}

In BB18 we reported a 3.5$\sigma$ detection of a source when
all three observations were stacked at the predicted position of Dysnomia, corresponding to
a body with a diameter of $700\pm 115$ and low albedo. Because of the unexpected nature of this result, the 
modest statistical significance, and the need to stack multiple datasets from different days to obtain a detection,
we re-observed the Eris-Dysnomia system with ALMA on 11 October 2018.  We obtained a single epoch in Band 6 (center frequency $\sim$233 GHz), in configuration C43-6 (resulting resolution $0.15\times0.12$ arcsec), with an on-source integration time of 78 minutes.  That observation should have sufficient sensitivity to detect a large Dysnomia, and sufficient resolution to separate it from Eris. We reduced the data in an identical
manner to that described in BB18, using the QSO J0006-0623 for pointing, atmosphere, bandpass, and flux density scale calibration, and the QSO J0141-0202 for complex gain as a function of time calibration.  The image resulting from these data is shown in Figure~\ref{fig:dysnomia}, where Dysnomia is clearly visible to the North and East of Eris, in its expected position.  We fit visibilities to a two-source model and find flux densities of $390\pm7 \mu$Jy for Eris and $45\pm7 \mu$Jy for Dysnomia.  That flux density for Dysnomia (a 6.5$\sigma$ detection) is close to that expected for a large Dysnomia, similar to what we found in BB18.  We note that the errors in fitting the flux density for Eris are reduced significantly when adding the second model source for Dysnomia (rather than having a single source for Eris), and that the final fitted positions for Eris and Dysnomia are insensitive to their positions in the initial model provided to the fit.  We add this Band 6 flux density of Dysnomia (and the simultaneously measured flux density of Eris) to our thermal model from BB18 and derive a new size for Dysnomia of 615$^{+60}_{-50}$km, with an albedo of 0.05$\pm$0.01. Figure~\ref{fig:thermal_model} shows an updated
version of Figure 7 of BB18 where we add the Band 6 flux densities for
Eris and Dysnomia at 1280 $\mu$m. 

For a system mass of $1.65\times10^{22}$ kg, the density of Dysnomia is $0.7\pm0.5$ g cm$^{-3}$
(1-$\sigma$
upper limit of 1.2 g cm$^{-3}$). The density of Eris is 2.4 g cm$^{-3}$ \citep{2021Icar..35514130H}. 
To high confidence, the density of Dysnomia
is significantly smaller than that of Eris.

{ Recent observations have found that the spin period 
of Eris is consistent with the orbital period of
Dysnomia, i.e. that Eris is phase locked to the orbit
of Dysnomia \citep{2023A&A...669L...3S, 2023arXiv230313445B}. \citet{2023A&A...669L...3S}
suggest that for tidal evolution to produce this
state within the age of the solar system
requires a Dysnomia with a density $>1.8$ g cm$^{-3}$,
a uniquely high value for any known
object this size. They predict a Dysnomia-Eris mass
ratio of $0.01-0.03$, which is between 1.4 and 7 $\sigma$ above our measured value. 
\citet{2023arXiv230313445B}, in contrast, considered tidal evolution of Dysnomia 
with a mass more consistent with our new measurement and concluded that Dysnomia
could become phase-locked if Eris were unusually dissipative. Such high dissipation
could have important consequences for the internal structure of Eris.
}
\section{Discussion}
The Vanth-Orcus mass ratio of 0.16$\pm$0.02 is larger
than that even of Charon-Pluto, which has a ratio of 0.12,
making the Vanth-Orcus mass ratio the highest of any known planet
or dwarf planet.
Like Pluto
and Charon, Orcus and Vanth have a similar density
(although the uncertainty on the Vanth density is
sufficiently large that future refinement
could change this conclusion) and lie on a circular
or nearly-circular orbit. 
Such a system is an expected outcome in the simulations of \citet{2019NatAs...3..802A} for either
differentiated or undifferentiated bodies impacting at near the escape velocity at
an impact angle greater than about 45$^\circ$. In this scenario, Vanth would be a nearly-intact
body, having lost little mass in the collision. The separation of Orcus and Vanth is close to that 
expected for creation through a giant impact and full tidal evolution to a double synchronous
state \citep{2021psnh.book..475C}. It appears likely that, like Pluto-Charon, Orcus-Vanth has achieved
this state.
The major difference between the two systems is
the apparent lack of system of small satellites
outside the orbit of Vanth, though satellites with the same fractional brightness as
the small ones of Pluto are still beyond the observational limits of any search
\citep{2008ssbn.book..335B}.

The Eris-Dysnomia system, with an upper limit to the mass ratio of 0.0085,
lies close to the transition region in \citet{2019NatAs...3..802A} between low mass ratio 
satellites formed out of a reaccreted disk material and a small intact fragment.
In both cases, the moon is predicted to have an ice fraction near 100\%,
consistent with the very low density of Dysnomia. An important clue is
perhaps the low albedo of Dysnomia. If the very low mass ratio small satellites of Pluto
and of Haumea can be used as representatives of disk reaccretion -- which is far from 
certain -- their high albedos are perhaps the signature of processing through a disk 
and removal of whatever volatile materials lead to space weathered darkening. 
A large intact fragment could retain its complement of darkening material and
lead to the typical low albedo that Dysnomia appears to have. Understanding whether or not
this process
occurs requires a significantly greater understanding of icy disk processing and the
causes of the low albedos of the Kuiper belt. The Eris-Dysnomia system is in a range
of parameter space poorly sampled in the \citet{2019NatAs...3..802A} simulations, so more insight
could also be gained through attempting to simulate this system, specifically.

Giant impact appears the most likely formation for
these dwarf planet satellite systems, but 
inconsistencies in the picture remain. No model 
has successfully generated Pluto's small moon system
at its current distance from the primary \citep{2021psnh.book..475C}, Haumea and 
the low mutual velocity of its 
collisional family remain unexplained, and the near-100\%
occurrence of satellites to these dwarf planets is
surprising. Dwarf planet satellite systems yield unique
insights into early solar system history and icy collisional
physics and continued study will provide an important window
into these processes.

\acknowledgements
This paper makes use of ALMA data: ADS/JAO.ALMA\#2018.1.00929.S, ADS/JAO.ALMA\#2015.1.00810.S, and  ADS/JAO.ALMA\#2016.1.00830.S.  ALMA is a partnership of ESO (representing its member states), NSF (USA) and NINS (Japan), together with NRC (Canada), MOST and ASIAA (Taiwan), and KASI (Republic of Korea), in cooperation with the Republic of Chile. The Joint ALMA Observatory is operated by ESO, AUI/NRAO and NAOJ. The National Radio Astronomy Observatory is a facility of the National Science Foundation operated under cooperative agreement by Associated Universities, Inc

\bibliography{alma}
\bibliographystyle{aasjournal}

\begin{turnpage}
\begin{deluxetable}{lcccccccccccc}
\renewcommand\thetable{1}

\tablecaption{ \label{astrometry}
   Astrometric positions and errors (milliarcseconds).
   }
\tablehead{
\colhead{}&\colhead{} &\colhead{} &\colhead{} &\colhead{} & \colhead{Right Ascension} &\colhead{} &\colhead{} & \colhead{}& \colhead{}& \colhead{Declination} & \colhead{}& \\
\colhead{Bodies} & \colhead{Date} & \colhead{Beam\tablenotemark{a}} & \colhead{Offset} & \colhead{ff error\tablenotemark{b}} & \colhead{sf error\tablenotemark{c}} & \colhead{cf error\tablenotemark{d}} & 
\colhead{total error} & \colhead{Offset} & \colhead{ff error} & \colhead{sf error} & \colhead{cf error} & 
\colhead{total error}} 
\startdata
Eris  & 2015-Nov-09 &  17 X  15 @ 78$^\circ$ &  -82.5 &  1.5 &  1.8 &   0.1 &  2.4 & -123.8 &  1.5 &   0.2 &  0.2 &  1.5 \\
Eris  & 2015-Nov-13 &  22 X  14 @ 51$^\circ$ &  -88.8 &  1.5 &  1.5 &   0.9 &  2.3 & -121.6 &  1.5 &   0.7 &  0.3 &  1.6 \\
Eris  & 2015-Dec-04 &  37 X  22 @ 63$^\circ$ &  -90.9 &  2.3 &  1.1 &   0.1 &  2.6 & -120.8 &  2.2 &   0.8 &  1.8 &  3.0 \\
Orcus & 2016-Oct-11 &  98 X  90 @ -56$^\circ$ &  -43.5 &  1.7 &  0.4 &   2.8 &  3.3 &  -24.0 &  1.6 &   2.7 &  4.0 &  5.1 \\
Vanth & 2016-Oct-11 &     "     & +145.4 &  6.6 &  7.7 &   2.8 & 10.5 & +137.9 &  6.2 &   0.3 &  4.0 &  7.4 \\
Orcus & 2016-Oct-13 & 107 X  93 @ -47$^\circ$ &  -26.8 &  3.7 &  9.9 &  32.3 & 34.0 &  +25.3 &  3.5 &   2.1 & 36.0 & 36.2 \\
Vanth & 2016-Oct-13 &     "    &  +178.2 & 14.4 &  2.2 &  32.3 & 35.4 & -101.6 & 14.0 &  27.2 & 36.0 & 47.3 \\
Orcus & 2016-Oct-15 & 123 X 103 @ 69$^\circ$ &  +16.3 &  1.8 &  0.4 &   7.8 &  8.0 &  +22.1 &  1.7 &   1.6 &  4.0 &  4.6 \\
Vanth & 2016-Oct-15 &     "     &  -96.6 &  5.8 &  2.7 &   7.8 & 10.1 & -205.1 &  5.5 &   5.8 &  4.0 &  8.9 \\
Orcus & 2016-Nov-07 & 197 X 163 @ 55$^\circ$ &   -2.6 &  2.3 &  6.7 &  16.5 & 18.0 &  -24.6 &  2.3 &   7.7 & 20.0 & 21.6 \\
Vanth & 2016-Nov-07 &     "     &  -60.1 &  6.8 &  2.1 &  16.5 & 18.0 & +165.6 &  6.7 &  17.2 & 20.0 & 27.2 \\
\enddata
\tablenotetext{a}{Synthesized beam FWHM axes and position angle (North through East, or CCW) with a robust weighting parameter of 0.}
\tablenotemark{b}{Formal error from fitting visibilities.}
\tablenotemark{c}{Systematic fitting error.}
\tablenotemark{d}{Celestial frame error.}
\end{deluxetable}
\clearpage
\end{turnpage}

\end{document}